# The Science Case for a Titan Flagship-class Orbiter with Probes

Authors:
**Conor A. Nixon**, *NASA Goddard Space Flight Center, USA*
*Planetary Systems Laboratory, 8800 Greenbelt Road, Greenbelt, MD 20771*
(301) 286-6757  `conor.a.nixon@nasa.gov`
**James Abshire**, *University of Maryland, USA*
**Andrew Ashton**, *Woods Hole Oceanographic Institution, USA*
**Jason W. Barnes,** *University of Idaho, USA*
**Nathalie Carrasco,** *Université Paris-Saclay, France,*
**Mathieu Choukroun,** *Jet Propulsion Laboratory, Caltech, USA*
**Athena Coustenis,** *Paris Observatory, CNRS, PSL, France*
**Louis-Alexandre Couston**, *British Antarctic Survey, UK*
**Niklas Edberg,** *Swedish Institute of Space Physics, Sweden*
**Alexander Gagnon,** *University of Washington, USA*
**Jason D. Hofgartner,** *Jet Propulsion Laboratory, Caltech, USA*
**Luciano Iess,** *University of Rome "La Sapienza", Italy*
**Stéphane Le Mouélic**, *CNRS, University of Nantes, France*
**Rosaly Lopes,** *Jet Propulsion Laboratory, Caltech, USA*
**Juan Lora,** *Yale University, USA*
**Ralph D. Lorenz**, *Applied Physics Laboratory, Johns Hopkins University, USA*
**Adrienn Luspay-Kuti,** *Applied Physics Laboratory, Johns Hopkins University USA,*
**Michael Malaska,** *Jet Propulsion Laboratory, Caltech, USA*
**Kathleen Mandt,** *Applied Physics Laboratory, Johns Hopkins University, USA*
**Marco Mastrogiuseppe,** *University of Rome "La Sapienza", Italy*
**Erwan Mazarico,** *NASA Goddard Space Flight Center, USA*
**Marc Neveu**, *University of Maryland, USA*
**Taylor Perron**, *Massachusetts Institute of Technology, USA*
**Jani Radebaugh**, *Brigham Young University, USA*
**Sébastien Rodriguez**, *Université de Paris, IPGP, Diderot, France*
**Farid Salama**, *NASA Ames Research Center, USA*
**Ashley Schoenfeld**, *University of California Los Angeles, USA*
**Jason M. Soderblom**, *Massachusetts Institute of Technology, USA*
**Anezina Solomonidou,** *European Space Agency/ESAC, Spain*
**Darci Snowden,** *Central Washington University, USA*
**Xioali Sun,** *NASA Goddard Space Flight Center, USA*
**Nicholas Teanby,** *School of Earth Sciences, University of Bristol, UK*
**Gabriel Tobie,** *Université de Nantes, France*
**Melissa G. Trainer,** *NASA Goddard Space Flight Center, USA*
**Orenthal J. Tucker,** *NASA Goddard Space Flight Center, USA*
**Elizabeth P. Turtle,** *Johns Hopkins Applied Physics Laboratory, USA*
**Sandrine Vinatier**, *Paris Observatory, CNRS, PSL, France*
**Véronique Vuitton,** *Université Grenoble Alpes, France*
**Xi Zhang,** *University of California Santa Cruz, USA*

Endorsed by:
**Veronica Allen**, *Universities Space Research Association, USA*
**Carrie Anderson**, *NASA Goddard Space Flight Center, USA*
**Shiblee Barua**, *Universities Space Research Association, USA*
**J. Michael Battalio**, *Yale University, USA*
**Patricia Beauchamp**, *Jet Propulsion Laboratory, California Institute of Technology, USA*
**Ross A. Beyer**, *SETI Institute and NASA Ames Research Center, USA*
**Julie Brisset**, *University of Central Florida, USA*
**John Cooper**, *NASA Goddard Space Flight Center, USA*
**Daniel Cordier**, *CNRS, Université de Reims, France*
**Martin Cordiner**, *Catholic University of America, USA*
**Thomas Cornet**, *Aurora Technology BV for ESA, ESAC, Spain*
**Ellen C. Czaplinski**, *University of Arkansas, USA*
**Chloe Daudon**, *Université de Paris, IPGP, France*
**Ravindra T. Desai**, *Imperial College London, UK*
**Shawn Domagal-Goldman**, *NASA Goddard Space Flight Center, USA*
**Chuanfei Dong**, *Princeton University, USA*
**Sarah Fagents**, *University of Hawaiʻi, USA*
**Tom G Farr**, *NASA Jet Propulsion Laboratory, Caltech, USA*
**William Farrell**, *NASA Goddard Space Flight Center, USA*
**Lori K. Fenton,** *SETI Institute, USA*
**Matthew Fillingim**, *University of California, Berkeley, USA*
**Timothy A. Goudge**, *The University of Texas at Austin, USA*
**Mark A Gurwell**, *Center for Astrophysics | Harvard & Smithsonian, USA*
**Jennifer Hanley**, *Lowell Observatory, USA*
**Tilak Hewagama**, *University of Maryland, USA*
**Amy E. Hofmann**, *NASA Jet Propulsion Laboratory, Caltech, USA*
**Sarah Hörst**, *Johns Hopkins University, USA*
**Der-You Kao,** *Universities Space Research Association, USA*
**Mathieu Lapôtre**, *Stanford University, USA*
**Sébastien Lebonnois**, *CNRS, Sorbonne Univ., France*
**Liliana Lefticariu**, *Southern Illinois University, USA*
**Alice Le Gall**, *LATMOS/IPSL, UVSQ, IUF, France*
**Liming Li**, *University of Houston, USA*
**Nicholas A. Lombardo**, *Yale University, USA*
**Chris McKay**, *NASA Ames Research Center, USA*
**Delphine Nna-Mvondo**, *University of Maryland Baltimore County, USA*
**Alena Probst**, *NASA Jet Propulsion Laboratory, CalTech, USA*
**Kerry Ramirez**, *Arizona State University, USA*
**Miriam Rengel**, *Max-Planck-Institut für Sonnensystemforschung, Germany*
**Emilie Royer**, *Planetary Science Institute, USA*
**Lauren Schurmeier**, *University of Hawaiʻi, USA*
**Edward Sittler**, *NASA Goddard Space Flight Center, USA*
**Jennifer Stern**, *NASA Goddard Space Flight Center, USA*

**Cyril Szopa**, *Université Paris-Saclay, France*
**Alexander Thelen**, *Universities Space Research Association, USA*
**Tuan H. Vu**, *NASA Jet Propulsion Laboratory, Caltech, USA*

## 1.0 Introduction – The Need for a Titan Orbiter with Probes

The joint NASA-ESA-ASI *Cassini-Huygens* mission [2, 3], which investigated the Saturnian system from 2004 to 2017, provided the first detailed look at its largest moon, Titan. Through 127 targeted flybys of the *Cassini* orbiter and *in situ* investigations of the lower atmosphere and surface by the *Huygens* probe, Titan was revealed for the first time. The mission uncovered surface features such as seas, lakes, dunes, mountains, filled and desiccated river valleys and plains [4-7] that were reminiscent of Earth in some respects but utterly different in others, as well as a dynamic atmosphere laden with organic molecules and replete with multiple layers of haze, clouds and rain [8-11] (Fig. 1). Titan provides a unique opportunity to study terrestrial processes in a completely different regime, and to learn about our home planet even as we learn about the solar system.

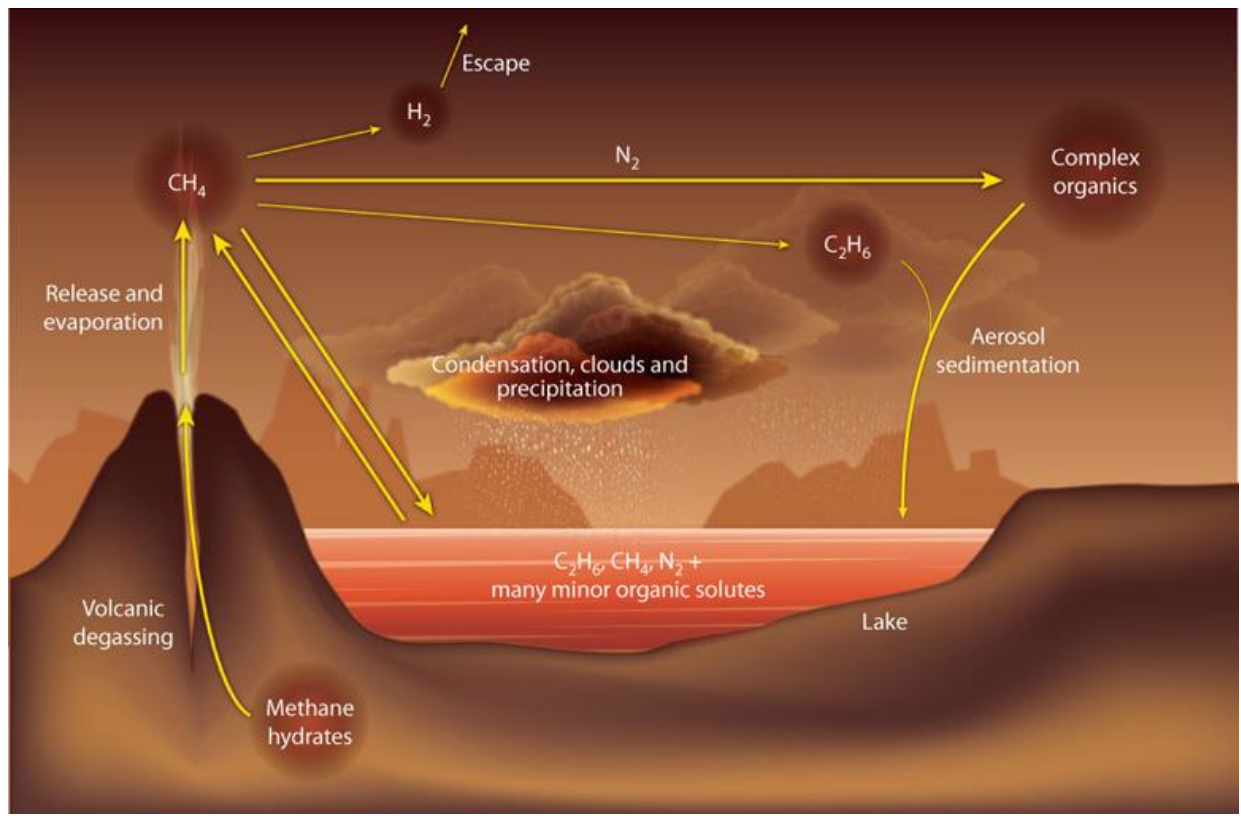

*Figure 1: Titan's atmosphere and seas are in constant interaction, through aerosol sedimentation, precipitation, evaporation and other processes. A flagship mission accomplishing remote sensing (orbiter) and in situ (probe) measurements, would build on* Cassini's *legacy, and provide insights to key questions both for Titan, and our solar system as a whole. (Figure credit: [1])*

In addition to its astounding successes, *Cassini-Huygens* left many questions about Titan unanswered (see Section 2). Insights into some of these questions will be provided by NASA's forthcoming *Dragonfly* mission, which will land on Titan's equatorial dune fields and sample the troposphere, surface and low latitude regions. However, *Dragonfly* will not address many other questions raised by *Cassini-Huygens* [12], including those related to global-scale geological history, atmospheric seasonal cycle and chemical processes, and origin and evolution of the polar seas, among other things. Furthermore, after Dragonfly only equatorial in situ measurements will have been obtained. Therefore, a complementary mission to investigate Titan's global-scale processes and make the first polar *in situ* measurements is highly desirable. These measurements could be accomplished by an instrumented orbiter, plus 1-2 entry probes that would investigate Titan's polar seas, giving the first ground-truth data from the only seas aside from Earth's found in the solar system, and a truly detailed global picture of Titan's dynamic atmosphere and surface.

Such a mission would advance NASA's most current goals in planetary science and should be pursued in the coming decade. For example, it would "advance scientific knowledge of the origin and history of the solar system, the potential for life elsewhere" (NASA Science Plan, 2020), and improve our understanding of "how geologic processes on Mars and on ocean-bearing worlds in our solar system might give rise to habitable environments". Similarly, it would address Strategic Objective 1.2: "scientific research of and from the Moon, lunar orbit, Mars, and beyond."

## 2.0 Titan from Cassini-Huygens

Through the *Cassini-Huygens* mission, Titan was discovered to have a subsurface water ocean, its surface was imaged in detail for the first time, and its atmospheric composition, thermal structure, and dynamics were examined from the surface to the exobase over two seasons.

At low latitudes, Titan was found to be mostly dry, except for large episodic methane rain storms [9] (Fig. 2(a)). The *Huygens* landing site was found to potentially be a dried river bed, with evidence of past streamflow. However, no liquids were observed filling the channels that were seen incising nearby hills [8] (Fig. 2(b)). Indeed Titan's methane hydrologic cycle is now the only other known example of such a cycle outside of Earth, composed of surface liquids, surface-atmosphere fluxes, atmospheric transport, and precipitation [13]. In addition, vast dune fields of apparently organic material girdle Titan's equator, reaching hundreds of meters in height and hundreds of kilometers in length [5] (Fig. 2(c)). At mid-latitudes, Titan exhibits a mostly bland, undifferentiated appearance of as-of-yet undetermined organic material [14], perhaps overlaying and masking any historical record of impacts or geologic activity in the buried regolith. Impact craters (Fig. 2(d)) were found to be less abundant than anticipated, indicating a youthful surface age of 200-1000 Myr [15]. The large craters that were seen provided evidence of erosional and possible tectonic activity [16-22]. Hence, Titan's surface seems to be dominated by sedimentary and/or depositional processes with dunes at the equator, plains at mid-latitudes, and dissected organic plateaux and lakes at the poles [7]. However, a full interpretation of the complex history, interaction, and ages of formation of the fluvial, aeolian, and impact features and processes **requires much greater compositional and morphological detail** from future missions.

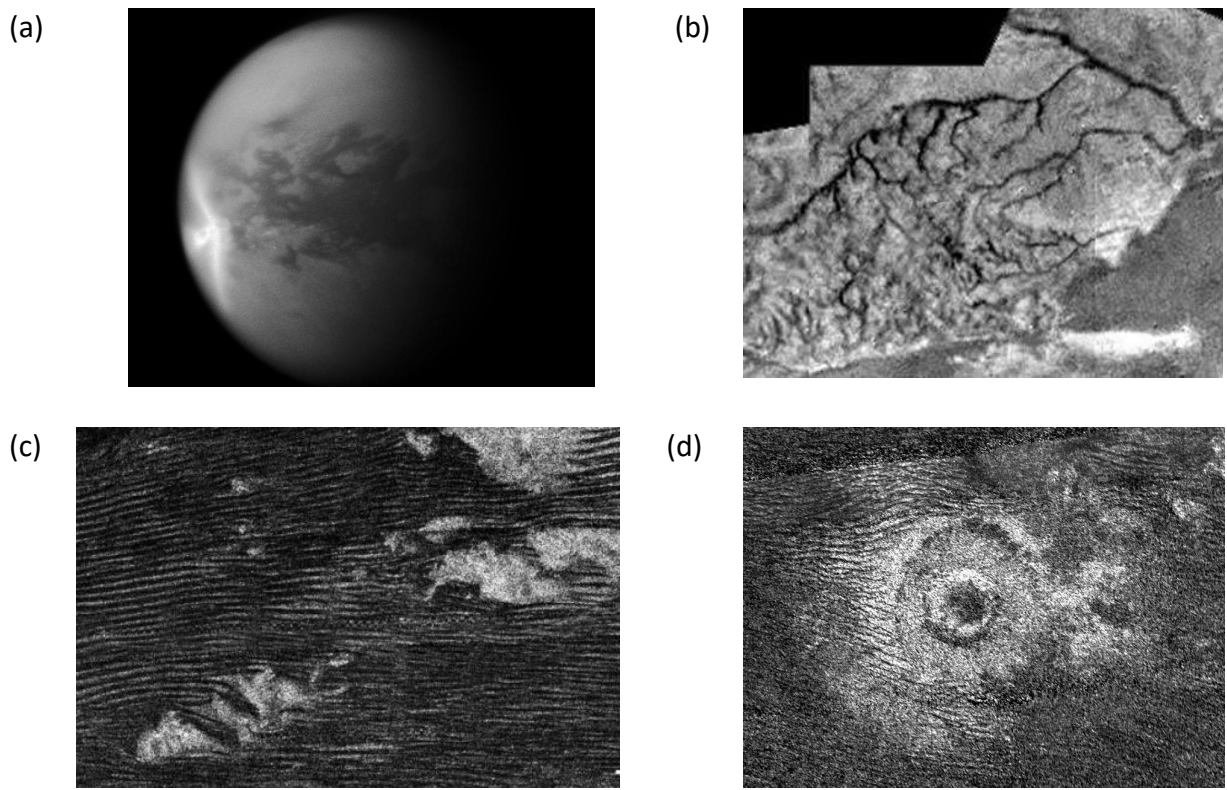

*Figure 2: Views of Titan at low latitudes. (a) a vast arrow-shaped storm sweeps around Titan, wetting the surface with methane rain (NASA/JPL/SSI); (b) branching river networks seen during the Huygens descent (NASA/ESA/University of Arizona); (c) equatorial dunes showing east-west zonal morphology (NASA/JPL); (d) dunes flowing around a crater (NASA/JPL).*

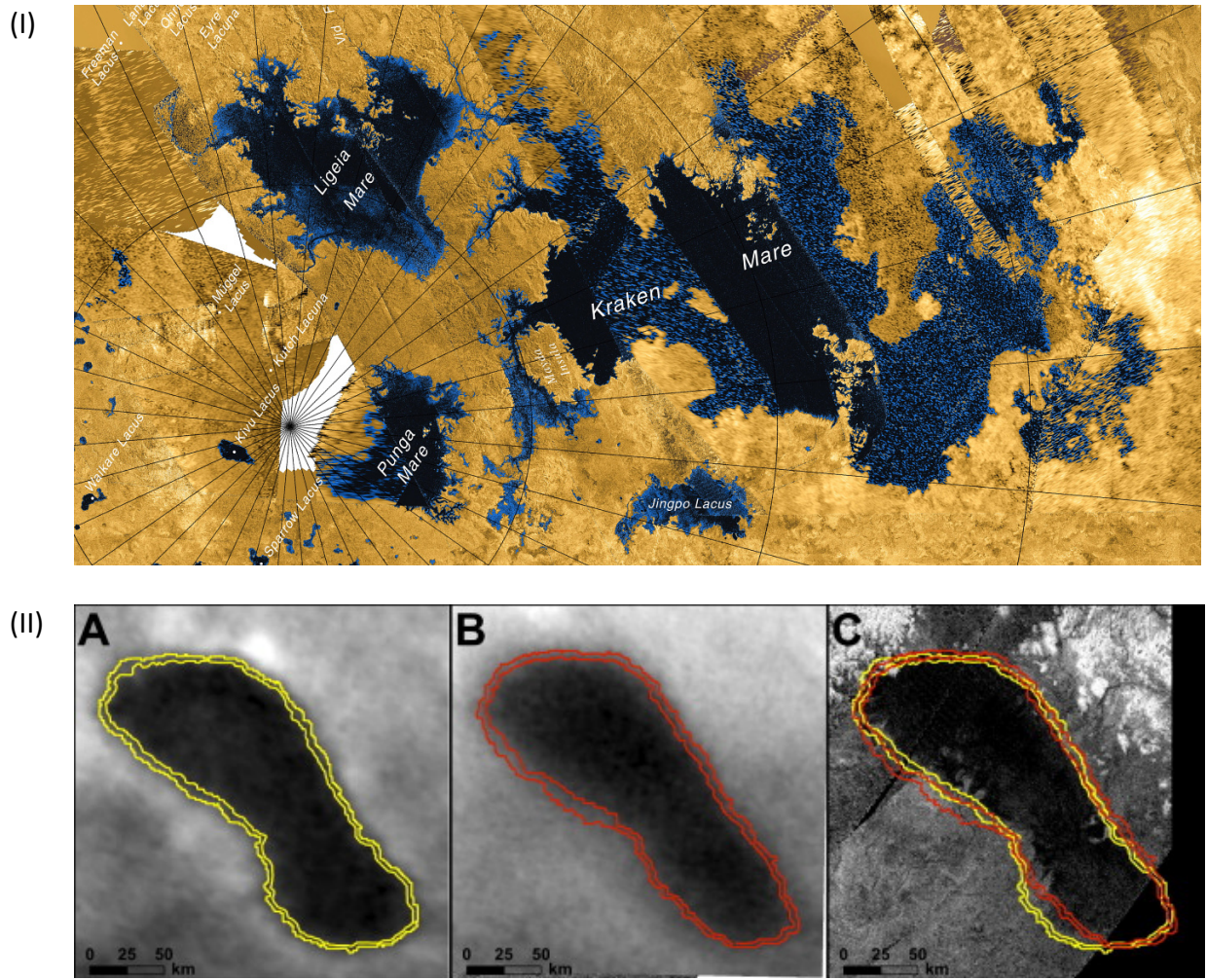

*Figure 3: Titan at high latitudes. (I) Northern seas (NASA/JPL/ASI); (II) Ontario Lacus near the south pole, showing changing shoreline from 2005 (A, from Cassini ISS) to 2009 (B, from ISS, and C, from Cassini radar). A polar orbiter could both map the polar regions at a consistent resolution and probe the sub-surface with radar, while probes could make measurements of the composition of the seas, giving insights into Titan's methane cycle and history.*

Titan's polar regions provided the biggest surprises of the mission. Large, partially interconnected seas and myriad small lakes of liquid hydrocarbons were discovered at the north pole [4]. *Cassini's* radar was able to determine bathymetric and crude compositional information for the largest seas [23] (Fig. 3(I)) such as Ligeia, showing depths reaching 160 m and a methane-rich composition. Titan's south pole appeared completely different, with only a single large lake (Ontario Lacus, Fig. 3(II)) with a composition including the less volatile hydrocarbon, ethane [24]. The great contrast between Titan's polar regions at the surface is only partially understood, but may in part be due to long-term cycles in Titan's seasons [25]. The recently-discovered presence of ramparts surrounding a few lakes imply that these lakes might be the youngest on Titan and require further investigation [26]. While there is **no direct evidence currently available about possible subsurface reservoirs,** they may have a large impact on climate stability [13, 27-29].

In the neutral atmosphere, *Cassini*'s instruments unveiled the distribution of organic compounds like never before, showing how molecules are transported from equator to pole in a global circulation that reverses direction annually [30, 31]. This reversal causes trace gases to seasonally accumulate over the winter pole and form ice clouds with complex compositions that need further investigation. The high-altitude 'detached' haze layer, first discovered by *Voyager 1* in 1980, was seen to migrate between hemispheres, rising and sinking in a complex interplay between chemistry and dynamics [10]. **However, the episodic nature of flyby observations have left a spatially and temporally incomplete picture of the seasonal changes occurring in Titan's atmosphere,** greatly limiting our ability to constrain climate models.

Above the neutral atmosphere, *Cassini*'s suite of six particle and field instruments also made unanticipated discoveries. The Ion and Neutral Mass Spectrometer (INMS) saw a rich cornucopia of organic molecular ions filling the mass range up to 100 Da [32]; the Cassini Plasma Spectrometer (CAPS) extended that range to hundreds of Da for positive ions and thousands of Da for negative ions, showing a continuum of masses from molecules to small aerosols [33]. The formation mechanisms, chemical composition, and evolution of these aerosols as they precipitate and travel through the atmosphere **remains uncertain, as there are many gaps in our understanding of their transition from molecules to particles.**

*Cassini* did not detect an internal magnetic field at Titan, suggesting a maximum permanent dipole moment of 0.78 nT $\times R_T^3$ [34]. The magnetic interaction between Titan's atmosphere and Saturn's outer magnetosphere is variable due to periodic flapping of Saturn's plasma sheet [35]. Cold, dense nitrile and hydrocarbon ions were observed above the exobase (~2000 to 3000 km altitude) and several Titan radii down the magnetospheric wake with estimates of mass loss rates due to ion outflow of $10^{27}$ amu/s [36-39]. This rate is significantly smaller than predictions of neutral loss rates near the exobase (~$10^{28}$ amu/s), leaving uncertainties about the atmospheric loss rates of $CH_4$ in particular [40] – **another major missing puzzle piece in our understanding of Titan's long-term atmospheric evolution and stability.**

*Cassini*'s Radio Science Sub-system (RSS) also probed Titan's interior, giving the first definitive confirmation that a deep water ocean exists [41, 42], inspiring interest and speculation in Titan's astrobiological potential. **However, the depth and boundaries of the internal water ocean have substantial uncertainties, requiring the collection of further data.**

Although a resounding success, the *Cassini-Huygens* mission raised as many questions about Titan as it answered [12, 43], including:
- What are the dominant escape processes at Titan's exobase, how do they vary with Titan's magnetospheric environment, and what is the fate of escaping molecules?
- What is the nature of the Saturn magnetosphere-Titan ionosphere interaction and which plasma processes contribute to atmospheric escape?
- Is the atmosphere in a steady-state, with on-going methane replenishment, or will it suffer long-term changes or collapse after methane depletion?
- What chemical pathways synthesize complex organic molecules in Titan's atmosphere?
- What chemistry and microphysics produce organic aerosols and clouds in the atmosphere?
- How fast do erosive processes act to obliterate surfaces features, including impact craters?
- How symmetric are the physical and chemical responses to Titan's seasons, and do liquids eventually migrate between hemispheres on long timescales leaving a climate record?
- What is the variability of composition between the lakes and seas, as well as surface terrains?
- Is there a vast, subsurface network of 'alkanofers' connecting the seas, acting as a reservoir for atmospheric methane?
- Does Titan have any internal geophysical activity (tectonic, seismic, cryovolcanic)?
- Do organic compounds from the atmosphere enter the subsurface ocean, producing a potentially habitable environment?

### 3.0 Titan Flagship Mission Concept

To fully understand Titan as a global system, a mission that addresses polar and global science questions is required. Key goals for such a mission include:

*(a) global, multi-spectral surface and subsurface mapping at a uniform resolution sufficient to identify geologic processes;*
*(b) temporal measurements of the atmosphere and exosphere to enable investigations of ongoing processes and seasonal change in the atmosphere and on the surface;*
*(c) in situ measurements of the polar regions, including the polar atmosphere and surface.*

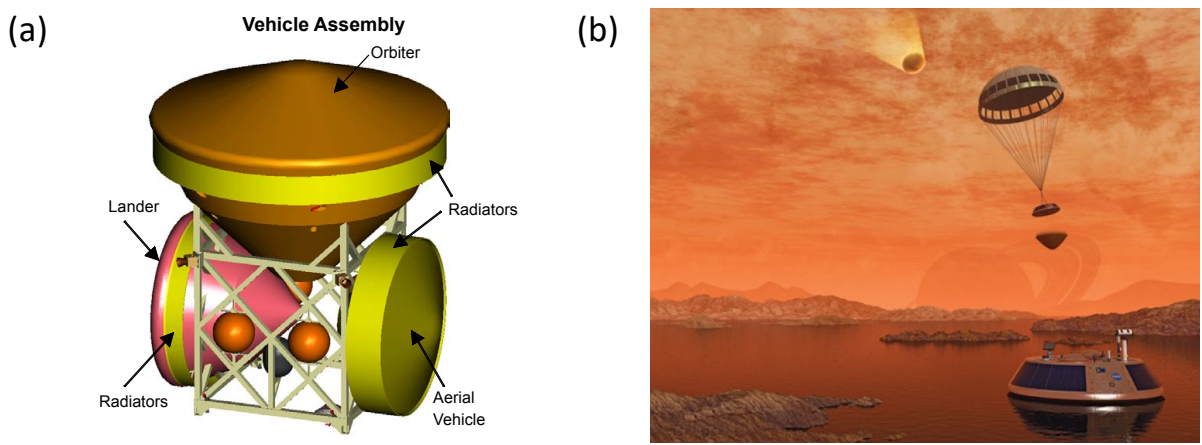

*Figure 4: (a) Titan orbiter, with lander and balloon components in cruise configuration, from the 2007 Titan flagship study. An orbiter with two sea-landing probes could be stowed in a similar configuration, but without the depicted aerocapture heatshield. (b) Artists concept of Titan Mare Explorer (TiME) Discovery-class mission (2011 proposal) (G. Murdoch, Pop. Sci.).*

The requirements for simultaneous global measurements and local polar information point to the need for dual investigation by a spacecraft in Titan polar orbit, surviving at least one full season (7.5 Earth years), and 1-2 probes targeting large polar seas and the near-surface environment. An orbiter with a suite of instruments would tackle the most important, high-level science questions raised by *Cassini-Huygens* at the global scale, while a probe would land in Kraken Mare, Titan's largest sea, and a possible second probe would target Ligeia Mare, Titan's second largest sea, to measure any compositional differences (see Fig. 4).

We envisage a flagship mission launching in the late 2020s. The final trajectory, including the gravity assist maneuvers, would be a factor determining launch vehicle size. It would be desirable for the mission to arrive at Saturn in the late 2030s, as the north pole emerges from winter and the probes are able to descend to the seas in daylight, with the possibility of direct radio monitoring from Earth. This would also allow direct comparison to *Voyager* and *Cassini* measurements made of the same season in 1980 and ~2010, two and one Titan years earlier, respectively.

Following cruise, the spacecraft would insert into Saturn's orbit, with apoapse near Titan. Probe release would occur during one of the subsequent Titan flybys, and the data relayed back to Earth. The probes for Titan's seas would be scaled down from *Huygens* or *TiME* [44] (Fig 4(b)). The probes would parachute to Titan's surface and use batteries for power, including thermal management, instrument operations and communications. Total operational lifetime would be several hours, sufficient to record data from the seas and relay back to the orbiter.

After probe data relay via the orbiter, the spacecraft would enter an elliptical polar orbit of Titan at high inclination (>80°), and later circularize using either atmospheric drag or a propulsive burn. The length of the circular mapping phase is expected to last at least 3 Earth years and would be determined during a mission design study. The orbiter vehicle would use next-generation RTGs for power and downlink the data to Earth via the Deep Space Network (DSN). The final instrument payload and number of probes would be determined by the available cost cap. An example payload for the orbiter and probes that meets the science requirements is given in Table 1.

*Table 1: Example Payload for Orbiter and Probes*

| Instrument | Science Driver |
| --- | --- |
| Orbiter | |
| Visible/NIR Color Camera | Take wide-angle, color images with multiple filters. |
| Vis/NIR Spectrometer | Make hyperspectral image cube maps of surface/atmosphere. |
| Topographic lidar | Topographic surface mapping for geology and geophysics. |
| Sub-mm sounder | Measure atmospheric temperature, composition and winds. |
| Surface-penetrating radar | Make 3D maps of Titan's subsurface. |
| Multistage mass spectrometer | Measure densities of major gases & ions. |
| Aerosol mass spectrometer | Particulate composition. |
| Magnetometer | Measure magnetic field environment. |
| Dual Langmuir probe | Upper atmosphere plasma fluid properties and E-field. |
| Gravity science investigation | Measure gravity field, atmospheric temperature. |
| Probes | |
| Atmospheric Structure Expt. | Measure temperature, pressure, hydrogen abundance. |
| Color Camera | Take color images of probe descent and from Titan's surface. |
| Surface Science Package | Measure winds, waves, dielectric constant, sea depth. |
| Mass spectrometer | Composition of the atmosphere, sea, and particulates. |

## 4.0 Summary and Conclusions

While our knowledge of Titan expanded dramatically during the *Cassini-Huygens* era, it has also greatly increased the scope of scientific questions to be addressed by future missions. *Cassini-Huygens* revealed that Titan is the only body with topography, meteorology, surface evolution, and atmospheric chemistry as rich and complex as Earth; however, we are limited in further inquiry by the restrictions of the current global dataset. **In addition, even after *Huygens* and *Dragonfly*, we will only have *in situ* data for low latitudes, whose nature strongly differs from the polar regions.** To address the large questions remaining after *Cassini-Huygens,* a new dedicated Titan mission that addresses both global and local (polar) science and provides long-term temporal coverage of the atmosphere is required. This points to the need for a flagship-class orbiter along with one or more probes targeting the northern seas. A mission study could be performed in the near future, including during the current decadal survey, to make a full costing and risk analysis of the strawman payload. Downsizing or changing instrument complement, as well as investigating opportunities for international partnerships, could be investigated. Such a study would provide a robust baseline for future decision-making and mission prioritization.

## References (uncondensed format)